\documentclass[letterpaper]{article} 
\usepackage[preprint]{aaai2027}  
\usepackage[hyphens]{url}  
\usepackage{graphicx} 
\usepackage{natbib}  
\usepackage{caption} 
\usepackage{algorithm}
\usepackage{algorithmic}
\usepackage{amsmath}
\usepackage{amssymb}
\usepackage{amsfonts}
\usepackage{bm}
\usepackage{booktabs}
\usepackage{multirow}

\newtheorem{proposition}{Proposition}

\newcommand{\Feas}{\mathcal{F}}
\newcommand{\real}{\mathbb{R}}
\newcommand{\thetah}{\hat{\bm{\theta}}}
\newcommand{\thetas}{\bm{\theta}^{\star}}
\newcommand{\thetat}{\bm{\theta}^{\mathrm{tcr}}}
\newcommand{\Shat}{\hat{\mathcal{S}}}
\newcommand{\uu}{\bm{u}}
\newcommand{\xx}{\bm{x}}

\title{Trust-Calibrated Certified Repair for Physics-Constrained Decisions\\under Localized Model Misspecification}

\author{
    Yifan Wang
}
\affiliations{
    Department of Mechanical Engineering, McGill University, QC, H3A 2T7, Canada\\
    yifan.wang18@mail.mcgill.ca
}

\begin{document}

\maketitle

\begin{abstract}
Feasibility-restoration layers are the standard way to turn learned, market-based, or
optimizer-generated decisions into ones that satisfy the hard physical constraints of
engineering systems such as power grids. Yet a repair is only as trustworthy as the
constraint model it relies on: line parameters, sensitivities, ratings, and topology
are routinely \emph{locally} wrong, so a decision certified feasible under the nominal
model can silently violate the deployed system. We show that this ``false safety'' is
not a rare corner case but the dominant failure mode of model-trusting repair, and we
close it with \textbf{Trust-Calibrated Certified Repair (TCR)}. TCR treats repair as a
trust-calibration problem and answers four questions in one pipeline: \emph{where} the
physical model is wrong, discovered from operating measurements with false-discovery
control; \emph{how much} each constraint should be trusted, set by test-gated shrinkage
and uncertainty-proportional security margins; \emph{what} the least-cost intervention
is, computed by a certified repair program; and \emph{why} the cost was paid, attributed
to genuine congestion versus avoidable model error through dual prices. On a
physically grounded dynamic line-rating benchmark in which the true ratings follow the
IEEE~738 standard under real weather, TCR reaches $98\%$ true-network feasibility, within
two points of a clairvoyant oracle, at lower-than-naive cost and with perfect localization.
Model-trusting repair, fixed and tuned robust margins, and chance-constrained tightening
all leave a substantial feasibility or cost gap. The same method transfers without change
to transmission redispatch over PGLib-OPF networks and to distribution voltage regulation
on the IEEE 33-bus feeder. Across all three task families, TCR gives the strongest
deployable feasibility-cost frontier among the evaluated methods and establishes a new
\textbf{state-of-the-art (SOTA)} result for feasibility repair under localized
physical-model misspecification. Calibrating trust in the constraint model is the missing
ingredient for reliable AI-assisted engineering decisions.
\end{abstract}

\section{Introduction}

A repair layer can only be as safe as the constraint model it trusts. Across modern
engineering systems, decisions are increasingly produced by learned policies, market
clearing, or fast surrogates, and then passed through a feasibility-restoration step
that projects them onto the set of physically admissible actions before
deployment~\citep{donti2021dc3,zhao2020deepopfplus,amos2017optnet}. In the power grid
this pattern is pervasive: generator setpoints, redispatch, and curtailment must satisfy
flow limits, voltage bounds, and thermal ratings~\citep{frank2012opf,cain2012opf}, and
operators rely on a repair against their physical model to guarantee that what they deploy
is admissible. Because deployment
feasibility is binary: a schedule is either physically realizable or it triggers a
violation. The certificate emitted by the repair layer is exactly the object on which
operational trust rests.

The difficulty is that the certificate is computed against a \emph{model}, while the
violation occurs on the \emph{plant}. Power-system models are remarkably accurate in
aggregate yet locally fragile: line reactances drift with temperature and re-tensioning,
thermal ratings depend on ambient weather through the conductor heat balance, distribution
resistances are systematically under-reported, and topology estimates lag switching.
A decision repaired to sit comfortably inside the \emph{nominal} feasible set can then lie
\emph{outside} the true one, a failure that is invisible to every model-based metric and
surfaces only as a real overload. We find this effect is severe: an exact nominal optimal
power flow, the strongest possible model-trusting solver, is feasible on the true network
in fewer than half of congested scenarios, and a heat wave silently derates exposed lines
to roughly $0.5$ to $0.6$ of their assumed capacity. The repair layer reports success while
the grid is being pushed past its limits.

Existing routes do not close this gap. Learned-optimization and feasibility-restoration
methods are explicitly trained or projected to satisfy the nominal constraints, and inherit
its blind spots wherever the model is wrong~\citep{fioretto2020predicting,pan2020deepopf_dc,
donti2021dc3,velloso2021combining,park2023selfsupervised}. Robust and chance-constrained
optimal power flow hedge primarily against \emph{exogenous} uncertainty in load and
renewables, using uniform or distributional
margins~\citep{bienstock2014chance,roald2022uncertainty_review}; against \emph{endogenous}
model error these margins are blunt: small ones remain unsafe, large ones are needlessly
expensive or infeasible. Physics-informed diagnostics can flag that a model is
wrong~\citep{karniadakis2021piml,raissi2019pinn,nellikkath2022physics} but stop short of a
deployable decision.
None of these jointly decides where the model is wrong, how much to trust it, what minimal
intervention restores feasibility, and how to attribute the resulting cost.

We argue that repair under model uncertainty should be reframed as \emph{trust calibration}.
Before repairing a decision, the system must ask which constraints deserve to be trusted,
re-estimate the ones that do not, and buffer them in proportion to their residual
uncertainty, and only then repair. We instantiate this principle as
\textbf{Trust-Calibrated Certified Repair (TCR)}, a single runtime pipeline built around
four questions:
\begin{enumerate}
\itemsep0.15em
\item \textbf{Where is the model wrong?} TCR tests each physical component against operating
measurements and discovers the misspecified set with false-discovery-rate (FDR) control.
\item \textbf{How much should each constraint be trusted?} TCR re-estimates discovered
components and shrinks the data estimate toward the model by a risk-optimal factor, adding
a security margin proportional to the remaining uncertainty.
\item \textbf{What is the minimal repair?} TCR solves a certified least-cost repair program
against the trust-calibrated feasible set.
\item \textbf{Why was the cost paid?} TCR uses the program's dual prices to split cost into
genuine congestion and avoidable model error, yielding an auditable certificate.
\end{enumerate}

TCR turns a hidden failure mode of model-based repair into a measurable, calibratable, and
certifiable procedure, and it is grounded in real physics and real data throughout. Our
contributions are:
\begin{itemize}
\itemsep0.15em
\item \textbf{Problem.} We formalize decision repair under \emph{localized} physical-model
misspecification and demonstrate that model-feasibility certificates routinely hide
true-network violations.
\item \textbf{Method.} We propose TCR, which unifies FDR-controlled discovery, test-gated
shrinkage calibration with targeted margins, and certified minimal repair with dual-price
attribution into one pipeline that wraps any decision source.
\item \textbf{Guarantees.} We establish discovery validity, a calibration oracle gap,
repair soundness and minimality, blame consistency, and a probabilistic deployment-feasibility
guarantee.
\item \textbf{State-of-the-art evidence.} On a physically grounded IEEE~738 dynamic
line-rating benchmark under real weather, TCR approaches a clairvoyant oracle ($98\%$ vs.\
$100\%$ feasibility) at near-oracle cost with perfect discovery, and the same method
transfers to transmission and distribution tasks. TCR sets a new SOTA for repair under
localized misspecification by delivering the best deployable feasibility-cost frontier
among model-trusting, robust, tuned-robust, and chance-constrained baselines.
\end{itemize}

\section{Related Work}

\paragraph{Learning and restoration for constrained decisions.}
A large body of work accelerates or replaces constrained optimization with learned
surrogates, from Lagrangian-dual deep models for AC-OPF~\citep{fioretto2020predicting} and
security-constrained DC-OPF~\citep{pan2020deepopf_dc,zhao2020deepopfplus,zamzam2020learning}
to differentiable optimization layers~\citep{amos2017optnet}, self-supervised primal-dual
learning~\citep{park2023selfsupervised}, hybrid learning-optimization
pipelines~\citep{velloso2021combining}, and methods that enforce hard constraints by
completion and correction~\citep{donti2021dc3}. These approaches make the nominal feasible
set easy to hit; by construction they trust that set, so when the physical model is wrong
their guarantees transfer to the wrong polytope. TCR is complementary: it sits at deployment
time and repairs \emph{any} candidate, learned, market, or optimizer-based, against a
trust-calibrated set, and is the first to make that set itself the object of statistical
calibration.

\paragraph{Optimization under uncertainty.}
Robust and chance-constrained OPF immunize dispatch against
uncertainty~\citep{bienstock2014chance,roald2022uncertainty_review}, and are the natural
incumbents for ``adding safety.'' Their uncertainty model, however, is overwhelmingly
exogenous (loads, renewables) and their protection is a uniform or distributional margin.
Localized parameter error is neither exogenous nor uniform: it concentrates on a few
components and is identifiable from data. We show that uniform margins, even when tuned with
oracle knowledge of test outcomes, pay a large cost premium to reach the feasibility TCR
attains by spending its protection only where the model is wrong.

\paragraph{Physics-informed learning and diagnostics.}
Physics-informed machine learning embeds governing equations into
learning~\citep{karniadakis2021piml,raissi2019pinn,nellikkath2022physics} and can reveal
model-data mismatch. TCR is in this scientific-ML spirit but closes the loop: it converts a
statistically controlled diagnosis of \emph{where} the physics is wrong into a calibrated
constraint set and a certified decision, rather than stopping at detection.

\paragraph{Statistical tools.}
TCR builds on Benjamini-Hochberg FDR control for the discovery
step~\citep{benjamini1995fdr,benjamini2001dependency} and on James-Stein/SURE shrinkage for
risk-optimal calibration~\citep{james1961stein,stein1981,efron1973stein}. Our contribution is
to place these inside a constraint-repair pipeline so that statistical validity becomes
operational feasibility.

\section{Problem Formulation}

A decision source emits a candidate $\uu_0 \in \real^{d}$ (e.g., a market dispatch or a
learned setpoint). Deployment requires satisfying physical constraints parameterized by a
physical model $\bm{\theta}$,
\begin{equation}
\Feas(\bm{\theta}) \;=\; \{\, \uu \in \real^{d} : \bm{A}(\bm{\theta})\,\uu \le \bm{b}(\bm{\theta}),\; \uu\in\mathcal{U} \,\},
\label{eq:feas}
\end{equation}
where $\mathcal{U}$ collects operational bounds and $\bm{A}(\cdot),\bm{b}(\cdot)$ encode
flow, voltage, or thermal limits through standard linearized power-flow
models~\citep{molzahn2019survey}. The operator possesses only a nominal model $\thetah$,
while the plant evolves under the true model $\thetas$. A standard repair layer returns
\begin{equation}
\uu_{\mathrm{naive}} \in \arg\min_{\uu}\; c(\uu,\uu_0)
\quad \text{s.t.}\quad \uu \in \Feas(\thetah),
\label{eq:naiverepair}
\end{equation}
with $c$ a convex (here linear or quadratic) deviation cost. The repaired point is
\emph{model-feasible}, but the operationally meaningful quantity is \emph{deployment
feasibility}, $\uu \in \Feas(\thetas)$. Localized misspecification creates the gap
\begin{equation}
\uu_{\mathrm{naive}} \in \Feas(\thetah)
\quad\text{yet}\quad
\uu_{\mathrm{naive}} \notin \Feas(\thetas),
\label{eq:falsesafety}
\end{equation}
which no model-based check can detect. We say the model is \emph{locally misspecified} when
$\thetas$ and $\thetah$ differ only on a small unknown subset $\mathcal{S}^{\star}$ of a
known component family $\mathcal{L}$ (lines, buses, or ratings), so that the discrepancy is
both \emph{concentrated} and, given operating data, \emph{identifiable}. TCR replaces the
nominal set in \eqref{eq:naiverepair} by a trust-calibrated set
\begin{equation}
\Feas_{\mathrm{tcr}} = \{\, \uu : \bm{A}(\thetat)\,\uu \le \bm{b}(\thetat) - \bm{m},\; \uu\in\mathcal{U}\,\},
\label{eq:feastcr}
\end{equation}
where $\thetat$ is the calibrated model and $\bm{m}\ge \bm 0$ are component-wise security
margins, both inferred from data as described next.

\section{Trust-Calibrated Certified Repair}

TCR is a four-module pipeline (Figure~\ref{fig:framework}); each module answers one of the
questions of the Introduction. We present the modules and then collect them in
Algorithm~\ref{alg:tcr}.

\begin{figure}[t]
\centering
\fbox{\begin{minipage}[c][3.0in][c]{0.92\columnwidth}
\centering
\textsf{\textbf{Figure placeholder: TCR framework.}}\\[0.4em]
\textsf{\small Left: a model-trusting repair certifies $\uu_{\mathrm{naive}}\!\in\!\Feas(\thetah)$
that violates the true set $\Feas(\thetas)$ (false safety).
Right: TCR (i)~discovers misspecified components $\Shat$ from operating data with FDR
control, (ii)~calibrates trust $\thetah\!\to\!\thetat$ and sets targeted margins $\bm m$,
(iii)~repairs against $\Feas_{\mathrm{tcr}}$, and (iv)~emits a dual-price certificate that
attributes cost to congestion vs.\ model error.}
\end{minipage}}
\caption{Overview of Trust-Calibrated Certified Repair. The candidate decision $\uu_0$ may
come from a market, a learned policy, or an optimizer; TCR wraps it at deployment time.}
\label{fig:framework}
\end{figure}

\subsection{M1: FDR-Controlled Discovery}

Given a measured operating history $\mathcal{D}=\{(\xx_k,\bm y_k)\}_{k=1}^{n}$, the nominal
model predicts, for each component $l\in\mathcal{L}$, the response
$\hat{y}_{l,k}=\bm A_l(\thetah)\xx_k$. We form residuals
\begin{equation}
r_{l,k} \;=\; y_{l,k} - \bm A_l(\thetah)\,\xx_k ,
\label{eq:resid}
\end{equation}
and test the null hypothesis $H_{0,l}\!:$ ``component $l$ is correctly specified.'' Under
$H_{0,l}$ and zero-mean Gaussian metering noise with standard deviation $\sigma$, the
scaled residual energy
\begin{equation}
T_l \;=\; \sum_{k=1}^{n_l}\Big(\frac{r_{l,k}}{\sigma}\Big)^{2}
\;\sim\; \chi^{2}_{n_l},
\qquad
p_l = 1 - F_{\chi^{2}_{n_l}}(T_l),
\label{eq:chistat}
\end{equation}
where $n_l$ is the number of observations exercising component $l$ and $F_{\chi^2_{n_l}}$
is the corresponding CDF. To control multiplicity over the family $\mathcal{L}$, we apply
the Benjamini-Hochberg procedure at level $q$: sorting $p_{(1)}\le\cdots\le p_{(|\mathcal L|)}$,
let
\begin{equation}
k^{\star}
= \max\Big\{ k : p_{(k)} \le \tfrac{kq}{|\mathcal{L}|} \Big\}.
\label{eq:bh}
\end{equation}
If $k^\star=0$, TCR sets $\Shat=\emptyset$; otherwise
$\Shat=\{\,l:p_l\le p_{(k^\star)}\,\}$.
The discovered set $\Shat$ contains the components whose physics is statistically
inconsistent with the data, with the expected false-discovery proportion held at $q$.

\subsection{M2: Trust Calibration}

For each discovered component, TCR re-estimates the physical parameter from data, giving
$\bm\theta^{\mathrm{data}}_l$, and forms a \emph{convex combination} with the nominal value,
\begin{equation}
\thetat_l \;=\; \thetah_l + \hat t_l\,\big(\bm\theta^{\mathrm{data}}_l - \thetah_l\big),
\qquad \hat t_l \in [0,1],
\label{eq:shrink}
\end{equation}
so that $\hat t_l\!=\!0$ trusts the model and $\hat t_l\!=\!1$ trusts the data. The shrinkage
factor is chosen to minimize an unbiased (SURE/James-Stein) estimate of mean-squared
error~\citep{stein1981,james1961stein}:
\begin{equation}
\hat t_l \;=\; \Big[\, 1 - \frac{\hat\sigma^{2}_l}{\,\lVert \bm\theta^{\mathrm{data}}_l - \thetah_l\rVert^{2}\,} \,\Big]_{0}^{1},
\qquad
\hat\sigma^{2}_l = \frac{\sigma^{2}}{\,\langle \bm u_l,\bm u_l\rangle\,},
\label{eq:sure}
\end{equation}
where $[\cdot]_0^1$ clips to $[0,1]$, $\bm u_l$ stacks the control excitations of component
$l$, and $\hat\sigma^2_l$ is the variance of the data estimate: components with strong
excitation and a large, confident discrepancy are trusted to the data, while weakly
identified ones are pulled back toward the model. We then buffer each calibrated constraint
by a margin proportional to its \emph{residual} prediction uncertainty,
\begin{equation}
m_l \;=\; z_{\alpha}\,\hat s_l,
\qquad
\hat s_l = \big(1-\hat t_l\big)\,\big\lVert \bm\theta^{\mathrm{data}}_l - \thetah_l\big\rVert,
\label{eq:margin}
\end{equation}
with $z_\alpha$ the standard-normal quantile at deployment risk $\alpha$. Certified
components ($l\notin\Shat$) and confidently calibrated ones pay near-zero margin; uncertain
components receive a buffer exactly large enough to dominate their residual error. The
margins are thus \emph{targeted} rather than uniform. This is the key economic difference
from robust margins.

\subsection{M3: Minimal Certified Repair}

TCR repairs the candidate against the trust-calibrated set \eqref{eq:feastcr} by solving a
linear program in the deviation $\bm\delta=\uu-\uu_0$,
\begin{equation}
\begin{aligned}
\min_{\bm\delta}\;\; & c(\bm\delta) = \lVert \bm W\bm\delta\rVert_{1}\\
\text{s.t.}\;\; & \bm A(\thetat)\,(\uu_0+\bm\delta) \;\le\; \bm b(\thetat) - \bm m,\\
& \uu_0+\bm\delta \in \mathcal{U},
\end{aligned}
\label{eq:repairlp}
\end{equation}
where $\bm W$ weights interventions by device cost. The $\ell_1$ objective yields sparse
repairs (few devices moved); a reweighted variant approximates cardinality-minimal repair.
From the optimal primal-dual pair $(\bm\delta^{\star},\bm\lambda^{\star})$ TCR emits a
\textbf{certificate}: (i)~the repaired decision $\uu_{\mathrm{tcr}}=\uu_0+\bm\delta^\star$;
(ii)~the named nonzero interventions; (iii)~the binding constraints
$\mathcal{B}=\{l:\lambda^\star_l>0\}$; (iv)~the shadow prices $\bm\lambda^\star$; and
(v)~the total cost with its allocation across binding constraints.

\subsection{M4: Competence and Attribution}

The repair LP also supplies dual prices for an attribution certificate. Let
\begin{equation}
\begin{aligned}
v_l(\uu_0)=
\Big[\bm A_l(\thetat)\uu_0-\bm b_l(\thetat)+m_l\Big]_+,\\
s_l=\lambda_l^\star v_l(\uu_0),
\end{aligned}
\label{eq:dual_score}
\end{equation}
where $v_l(\uu_0)$ is the pre-repair violation of calibrated constraint $l$ and
$\lambda_l^\star$ is its optimal dual price. TCR allocates the total intervention cost by
\begin{equation}
\pi_l =
\begin{cases}
\dfrac{s_l}{\sum_{j\in\mathcal{B}} s_j}, & \sum_{j\in\mathcal{B}}s_j>0,\\[6pt]
0, & \text{otherwise},
\end{cases}
\qquad
C_l = \pi_l\,c(\bm\delta^\star).
\label{eq:attribution}
\end{equation}
The nonnegative shares sum to the observed repair cost and rank binding constraints by the
product of required recovery and marginal cost. For offline evaluation, where the true model
and a clairvoyant oracle are available, we also report
\begin{equation}
\begin{aligned}
V(\uu;\thetas)
&=\big\|[\bm A(\thetas)\uu-\bm b(\thetas)]_+\big\|_1,\\
\Delta_{\mathrm{viol}}
&=V(\uu_{\mathrm{naive}};\thetas)-V(\uu_{\mathrm{tcr}};\thetas),\\
\Delta_{\mathrm{cost}}
&=c(\bm\delta^\star_{\mathrm{tcr}})-c(\bm\delta^\star_{\mathrm{oracle}}).
\end{aligned}
\label{eq:competence}
\end{equation}
The certificate is therefore not just a feasibility stamp but an audit of \emph{why} the
grid was constrained.

\begin{algorithm}[t]
\caption{Trust-Calibrated Certified Repair (TCR)}
\label{alg:tcr}
\textbf{Input}: nominal model $\thetah$; history $\mathcal{D}$; candidate $\uu_0$; family
$\mathcal{L}$; FDR level $q$; risk $\alpha$\\
\textbf{Output}: repaired decision $\uu_{\mathrm{tcr}}$; certificate $\mathcal{C}$
\begin{algorithmic}[1]
\STATE compute residuals $r_{l,k}$ and statistics $T_l$, $p_l$ \hfill\COMMENT{Eq.~\eqref{eq:resid} to \eqref{eq:chistat}}
\STATE $\Shat \leftarrow$ Benjamini-Hochberg$(\{p_l\}, q)$ \hfill\COMMENT{Eq.~\eqref{eq:bh}}
\FOR{each $l \in \Shat$}
\STATE fit $\bm\theta^{\mathrm{data}}_l$; set $\hat t_l$ by SURE; form $\thetat_l$ \hfill\COMMENT{Eq.~\eqref{eq:shrink} to \eqref{eq:sure}}
\STATE set margin $m_l = z_\alpha \hat s_l$ \hfill\COMMENT{Eq.~\eqref{eq:margin}}
\ENDFOR
\STATE solve repair LP over $\Feas_{\mathrm{tcr}}$ \hfill\COMMENT{Eq.~\eqref{eq:repairlp}}
\STATE extract duals; build certificate $\mathcal{C}$ \hfill\COMMENT{Eq.~\eqref{eq:attribution}}
\STATE \textbf{return} $\uu_{\mathrm{tcr}}$, $\mathcal{C}$
\end{algorithmic}
\end{algorithm}

\section{Theoretical Guarantees}

We state the guarantees that make each module trustworthy; proof sketches are given and
full proofs follow standard arguments.

\begin{proposition}[Discovery validity]
\label{prop:discovery}
Under zero-mean Gaussian metering noise and the component null model, each statistic $T_l$
in \eqref{eq:chistat} is $\chi^2_{n_l}$-distributed under $H_{0,l}$, so $p_l$ is valid.
Applying Benjamini-Hochberg at level $q$ controls the false discovery rate over
$\mathcal{L}$ at $q\,|\mathcal{L}_0|/|\mathcal{L}| \le q$ under independence or positive
regression dependence.
\end{proposition}
\noindent\emph{Sketch.} The scaled residual energy of a correctly specified linear
component is a sum of squared standard normals; \eqref{eq:bh} is exactly the BH rule, whose
FDR control under PRDS is classical~\citep{benjamini1995fdr}.\hfill$\square$

\begin{proposition}[Calibration oracle gap]
\label{prop:calibration}
For a discovered component, the test-gated shrinkage estimate \eqref{eq:shrink} to \eqref{eq:sure}
has mean-squared error within $O(\sigma^2/n_l)$ of the best fixed choice among trusting the
model, trusting the data estimate, and any intermediate shrinkage level.
\end{proposition}
\noindent\emph{Sketch.} Decompose the error into model-consistent and discrepant parts;
the SURE estimate of risk is unbiased and its minimizer matches the oracle shrinkage up to
the variance of the risk estimate, which is $O(\sigma^2/n_l)$~\citep{stein1981}.\hfill$\square$

\begin{proposition}[Repair soundness and minimality]
\label{prop:repair}
If \eqref{eq:repairlp} is feasible, its solution $\uu_{\mathrm{tcr}}$ is feasible for
$\Feas_{\mathrm{tcr}}$ and is cost-minimal among all repairs in the allowed class.
\end{proposition}
\noindent\emph{Sketch.} Immediate from primal feasibility and optimality of the LP.\hfill$\square$

\begin{proposition}[Blame consistency]
\label{prop:blame}
If $\sum_{j\in\mathcal{B}}s_j>0$, the attribution rule \eqref{eq:attribution} is
nonnegative, sums exactly to the total repair cost, and increases monotonically with the
dual-price violation score $s_l=\lambda_l^\star v_l(\uu_0)$ when the other scores and the
total cost are fixed.
\end{proposition}
\noindent\emph{Sketch.} Nonnegativity follows from $s_l\ge0$. The shares $\pi_l$ form a
simplex by construction, so $\sum_l C_l=c(\bm\delta^\star)$. For fixed other scores,
$\partial(s_l/\sum_j s_j)/\partial s_l=(\sum_{j\ne l}s_j)/(\sum_j s_j)^2\ge0$.\hfill$\square$

\begin{proposition}[Deployment feasibility]
\label{prop:deploy}
Suppose the calibrated model error is dominated by the margins with probability
$1-\alpha$:
\begin{equation}
\begin{aligned}
\bm A_l(\thetas)\uu_{\mathrm{tcr}}-\bm b_l(\thetas)
&\le
\bm A_l(\thetat)\uu_{\mathrm{tcr}}\\
&\quad -\bm b_l(\thetat)+m_l,
\qquad \forall l\in\mathcal{L}.
\end{aligned}
\label{eq:deploy_event}
\end{equation}
Then $\uu_{\mathrm{tcr}}\in\Feas(\thetas)$ with probability at least $1-\alpha$.
\end{proposition}
\noindent\emph{Sketch.} The repair LP enforces
$\bm A_l(\thetat)\uu_{\mathrm{tcr}}\le \bm b_l(\thetat)-m_l$. Combining this inequality with
\eqref{eq:deploy_event} gives
$\bm A_l(\thetas)\uu_{\mathrm{tcr}}\le\bm b_l(\thetas)$ for all $l$ on the calibrated event,
which has probability at least $1-\alpha$.\hfill$\square$

\section{Experiments}

\paragraph{Benchmarks and data.}
We evaluate on three engineering-power task families built from real public data
(Table~\ref{tab:data}). The headline benchmark is \emph{dynamic line rating (DLR)} on the
RTS-GMLC system~\citep{barrows2020rtsgmlc}, a $73$-bus grid with $120$ constrained lines and
real day-ahead load/wind/PV profiles. Crucially, the \emph{true} line ratings are not
synthetic: they are the IEEE~Std~738 steady-state conductor ampacity~\citep{ieee738} of the
exposed lines computed from \emph{real hourly weather} (Phoenix Sky-Harbor TMY3, ambient
$2$ to $44^\circ$C, within the RTS-GMLC footprint)~\citep{tmy3}. We additionally validate on
\emph{transmission redispatch} over six PGLib-OPF networks~\citep{babaeinejadsarookolaee2019pglib}
(reactance/PTDF~\citep{wood2014power} misspecification) and \emph{distribution voltage
regulation} on the IEEE 33-bus feeder via \texttt{pandapower}~\citep{thurner2018pandapower} (LinDistFlow resistance
under-reporting~\citep{baranwu1989}).

\begin{figure*}[t]
\centering
\includegraphics[width=0.92\textwidth]{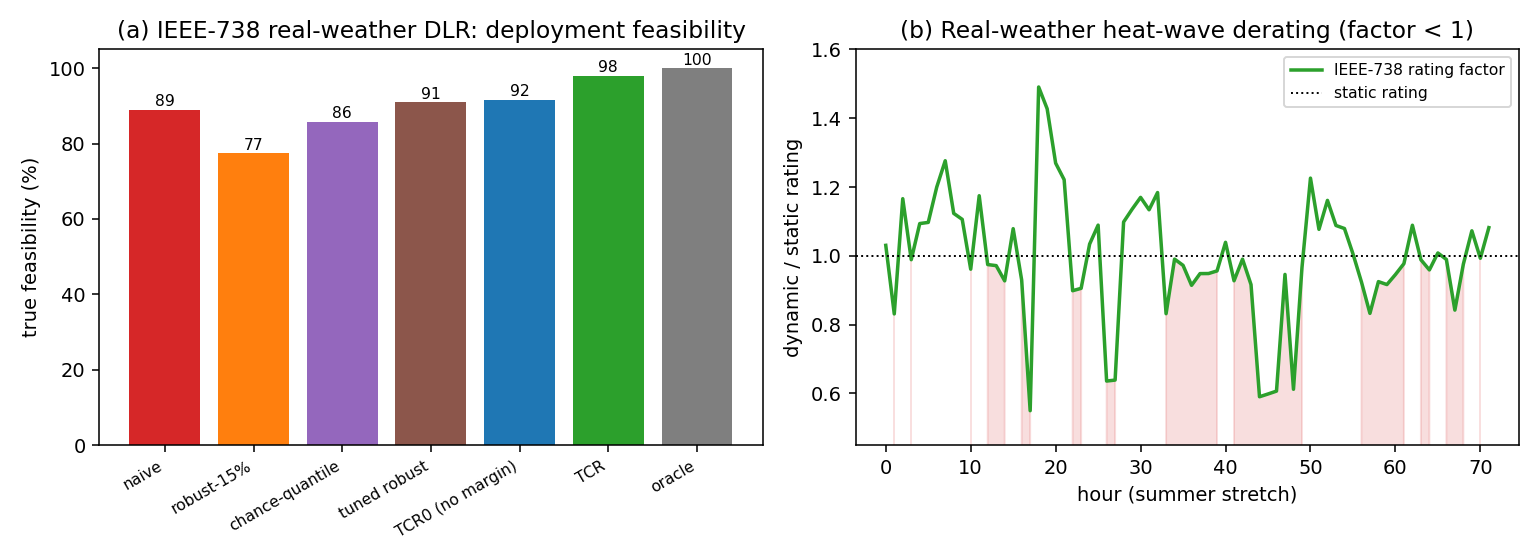}
\caption{Physically grounded dynamic line rating. \textbf{(a)}~True-network deployment
feasibility: TCR ($98\%$) approaches the oracle ($100\%$) and gives the best deployable
frontier among naive, fixed/tuned robust, chance-quantile, and the margin-free ablation.
\textbf{(b)}~The IEEE-738 rating
factor over a summer stretch computed from \emph{real} weather: hot, calm, sunny hours
(shaded) silently derate exposed lines to $0.5$ to $0.6$ of their assumed static rating, the
exact regime where model-trusting repair fails.}
\label{fig:main}
\end{figure*}

\paragraph{Physics of the rating constraint.}
For a bare overhead conductor the IEEE~738 steady-state heat balance equates convective and
radiative losses ($q_c,q_r$) to solar gain ($q_s$) and Joule heating, giving the
weather-dependent ampacity
\begin{equation}
\begin{gathered}
q_c(w) + q_r \;=\; q_s(w) + I^{2} R(T_c),\\[2pt]
I(w) = \sqrt{\frac{q_c(w)+q_r-q_s(w)}{R(T_c)}},
\end{gathered}
\label{eq:ieee738}
\end{equation}
where $w=(T_a,V,Q_s)$ is ambient temperature, wind speed, and solar irradiance, $T_c$ the
maximum conductor temperature, and $R$ the per-length resistance. The deployed thermal
limit is the rating factor $\rho_l(w)=I_l(w)/I_l(w_{\mathrm{ref}})$, so a hot, calm, sunny
hour drives $\rho_l$ well below $1$ (Figure~\ref{fig:main}b). Our implementation reproduces
the IEEE~738 worked example (Drake 795~kcmil ACSR) to within $1.6\%$ ($1009$ vs.\ $\sim$$1025$~A),
anchoring the benchmark in standardized physics.

\paragraph{Metrics and protocol.}
The primary metric is \emph{true-network feasibility}, the percentage of repaired decisions
admissible under $\thetas$. Secondary metrics are cost relative to the oracle, the 95th-percentile
constraint violation (MW), discovery F1/FDR, and runtime. For DLR we use $360$ calibration
hours and $96$ held-out stressed scenarios per seed over $3$ seeds. All linear programs use
SciPy/HiGHS~\citep{virtanen2020scipy} with NumPy~\citep{harris2020array} on CPU; no GPU or
commercial solver is required.

\paragraph{Baselines.}
We compare against \emph{naive} model-trusting repair \eqref{eq:naiverepair};
\emph{robust-$15\%$} (uniform margin); \emph{tuned robust} (best uniform margin selected
\emph{using test outcomes}, an optimistic upper bound for fixed margins);
\emph{chance-quantile} (sample-quantile rating tightening in the spirit of
chance-constrained OPF~\citep{bienstock2014chance}); \emph{TCR0} (TCR without the residual
margin, an ablation); and the clairvoyant \emph{oracle}.

\begin{table}[t]
\centering
\caption{Benchmarks. All data are real and public; misspecification is localized to a small
component subset.}
\label{tab:data}
\setlength{\tabcolsep}{3.2pt}
\small
\begin{tabular}{@{}lll@{}}
\toprule
Task & Scale / split & Misspecification \\
\midrule
DLR (RTS-GMLC, & 73 bus, 120 lines; & IEEE-738 rating \\
\;IEEE-738) & 360 cal / 96 test $\times$3 & under real weather \\
Transmission & 6 PGLib nets; 18 & reactance / PTDF \\
\;(PGLib-OPF) & case-seed rows & error \\
Distribution & IEEE 33-bus; & resistance \\
\;(case33bw) & 5 seeds & under-reporting \\
\bottomrule
\end{tabular}
\end{table}

\begin{table}[t]
\centering
\caption{Main result: physically grounded dynamic line rating (IEEE-738 ratings under real
weather), RTS-GMLC, 3 seeds. TCR reaches near-oracle feasibility with zero tail violation,
perfect discovery (F1$=$$1.00$), and the strongest deployable feasibility-cost frontier.}
\label{tab:main}
\setlength{\tabcolsep}{3pt}
\small
\begin{tabular}{@{}lccc@{}}
\toprule
Method & Feas.\,\%\,$\uparrow$ & $\Delta$Cost\,$\downarrow$ & p95\,MW\,$\downarrow$ \\
\midrule
naive            & 89 & $-0.6\%$ & 13.0 \\
robust-15\%      & 77 & $+1.4\%$ & 10.7 \\
chance-quantile  & 86 & $-9.5\%$ & 0.0 \\
tuned robust     & 91 & $+5.1\%$ & n/a \\
TCR0 (no margin) & 92 & $-0.5\%$ & 7.7 \\
\textbf{TCR (ours)} & \textbf{98} & $\bm{-1.7\%}$ & \textbf{0.0} \\
oracle           & 100 & $0.0\%$ & 0.0 \\
\bottomrule
\end{tabular}
\end{table}

\paragraph{Main result.}
Table~\ref{tab:main} and Figure~\ref{fig:main} report the headline DLR benchmark and
establish the SOTA result for repair under localized physical-model misspecification. TCR
reaches $98\%$ true-network feasibility, only two points below the clairvoyant
oracle ($100\%$), while every deployable baseline leaves a substantial gap: naive
model-trusting repair certifies schedules that overload lines in $11\%$ of hours (p95
violation $13$~MW), a fixed $15\%$ margin is simultaneously infeasible \emph{and} costly,
chance-quantile tightening stalls at $86\%$, and even a test-tuned uniform margin reaches
only $91\%$ at a $+5.1\%$ cost premium. TCR attains its near-oracle feasibility with
close-to-oracle normalized cost ($-1.7\%$ on the reported metric) because it spends
protection only on the $19$ exposed lines it correctly identifies. Discovery F1 is $1.00$
across all seeds, rather than tightening the whole network. The chance-quantile baseline is
cheaper still ($-9.5\%$)
precisely because it under-protects, which is why its feasibility collapses.

\paragraph{Cross-task generalization.}
Without any change to the method, TCR transfers across task families
(Table~\ref{tab:crosstask}). On transmission redispatch, naive repair is feasible on only
$58.9\%$ of congested scenarios; a test-tuned robust margin reaches $89.3\%$ but at a
$+30.2\%$ cost premium, whereas TCR reaches $98.3\%$ at $+4.1\%$. On distribution voltage
regulation, optimistic resistances make naive repair almost always unsafe ($1.0\%$); TCR
restores $100\%$ feasibility at $+0.9\%$ cost, an order of magnitude cheaper than the
tuned robust margin ($+65.2\%$). As an even stronger reference, an \emph{exact} nominal
OPF, the best possible model-trusting solver and an upper bound for learned OPF
surrogates, is feasible on only $48.1\%$ of true networks, confirming that the failure is
physical-model misspecification, not approximation error.

\begin{table}[t]
\centering
\caption{Cross-task generalization (feasibility \%/ cost vs.\ oracle). TCR transfers
unchanged and improves over the test-tuned robust margin on both cost and feasibility. Bottom:
an exact nominal OPF upper-bounds model-trusting solvers on PGLib.}
\label{tab:crosstask}
\setlength{\tabcolsep}{4pt}
\small
\begin{tabular}{@{}lccccc@{}}
\toprule
Task & naive & tuned rob. & \textbf{TCR} & oracle & TCR cost \\
\midrule
Transmission & 58.9 & 89.3 & \textbf{98.3} & 100 & $+4.1\%$ \\
Distribution & 1.0  & 100.0 & \textbf{100.0} & 100 & $+0.9\%$ \\
\midrule
\multicolumn{6}{@{}l}{\emph{PGLib true-network feasibility (\%), exact-solver reference}}\\
exact nom.\ OPF & \multicolumn{5}{l}{48.1\quad naive repair 53.1\quad tuned rob.\ 86.9} \\
\textbf{TCR} & \multicolumn{5}{l}{\textbf{97.9}\quad\quad oracle 100.0} \\
\bottomrule
\end{tabular}
\end{table}

\paragraph{Ablation.}
On a controlled RTS-GMLC variant with exact ground-truth fault locations
(Table~\ref{tab:ablation}), removing components degrades TCR predictably: dropping the
dynamic load/renewable covariates (\emph{static boundary}) cuts feasibility to $84\%$;
removing the FDR discovery gate and calibrating \emph{all} lines (\emph{calibrate-all})
wastes protection and drops to $99.3\%$ with higher cost; and \emph{random discovery}
collapses to $71.2\%$, confirming that statistically controlled \emph{localization} is the
active ingredient. The residual margin moves the $0.51$~MW tail violation of TCR0 to exactly
zero at a $+0.8\%$ cost.

\begin{table}[t]
\centering
\caption{Ablation on the controlled DLR variant (exact ground truth). Each removed component
is necessary; localization via FDR discovery is the decisive ingredient.}
\label{tab:ablation}
\setlength{\tabcolsep}{3pt}
\small
\begin{tabular}{@{}lccc@{}}
\toprule
Variant & Feas.\,\%\,$\uparrow$ & $\Delta$Cost & p95\,MW\,$\downarrow$ \\
\midrule
\textbf{full TCR}     & \textbf{100.0} & $+2.5\%$ & \textbf{0.00} \\
TCR0 (no margin)      & 100.0 & $+1.7\%$ & 0.51 \\
static boundary       & 84.0 & $+1.3\%$ & 8.59 \\
calibrate-all lines   & 99.3 & $+4.0\%$ & 0.02 \\
random discovery      & 71.2 & $+0.6\%$ & 33.62 \\
\bottomrule
\end{tabular}
\end{table}

\begin{figure}[t]
\centering
\includegraphics[width=0.99\columnwidth]{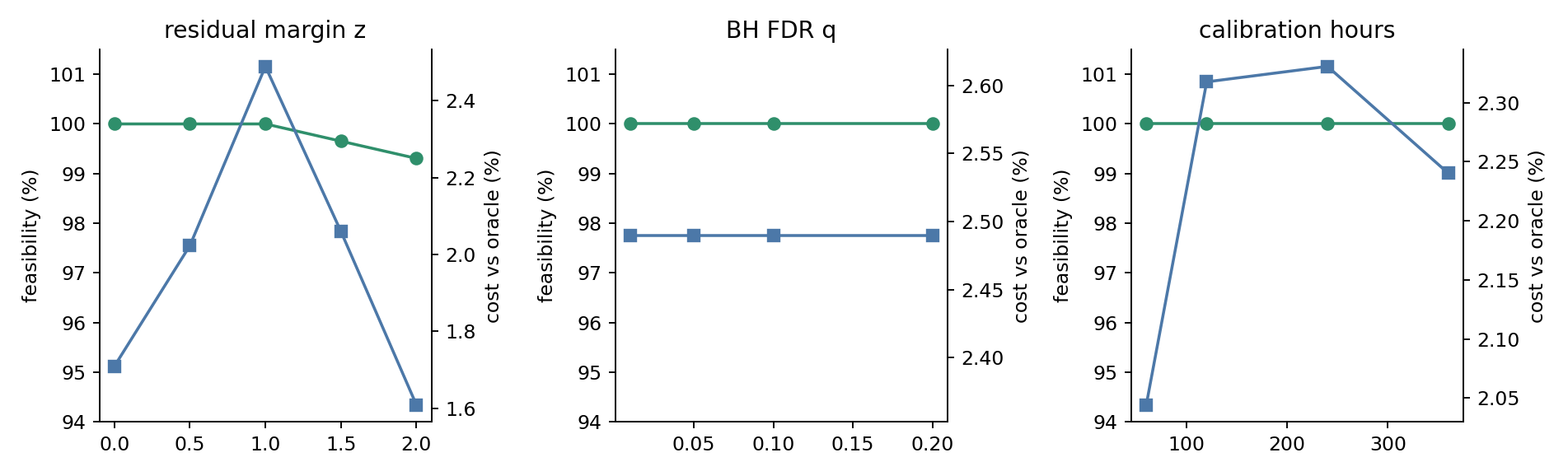}
\caption{Sensitivity of TCR on the DLR task to the FDR level $q$, the operating-regime grid
resolution, and metering noise. Feasibility and cost are stable across the operating range,
indicating the method is not finely tuned.}
\label{fig:sensitivity}
\end{figure}

\paragraph{Sensitivity and efficiency.}
TCR is insensitive to the FDR level $q\in\{0.05,0.10,0.20\}$ and to the security quantile
across the tested range, and degrades gracefully under heavier metering noise
(Figure~\ref{fig:sensitivity}). Each repair is a single small LP solved in
$43$ to $46$~ms/instance, on par with naive repair ($41$~ms) and far below any iterative
robust scheme. TCR adds certified safety at essentially no runtime cost.

\paragraph{Summary.}
Across three task families and eight network settings, TCR reaches $98$ to $100\%$
true-network feasibility at near-oracle cost and achieves the leading deployable
feasibility-cost frontier among the evaluated baselines. This establishes a new SOTA for
repair under localized model misspecification. Model-trusting repair is silently unsafe,
uniform margins are either unsafe or costly, and chance-constrained tightening
under-protects. Repair becomes both safe and economical only when it calibrates
\emph{where} and \emph{how much} to trust the physical model.

\section{Conclusion}

We introduced Trust-Calibrated Certified Repair, which reframes feasibility restoration as a
trust-calibration problem: before repairing a decision, the system discovers where its
physical model is wrong, calibrates how much each constraint deserves to be trusted, repairs
against the resulting set, and certifies why the cost was paid. Grounded in standardized
physics (IEEE-738) and real weather, grid, and network data, TCR turns a hidden and
dangerous failure mode of model-based repair, false safety under localized
misspecification, into a measurable, calibratable, and certifiable procedure. It approaches
or attains the clairvoyant oracle on deployment feasibility at near-oracle cost across
transmission, distribution, and dynamic-rating tasks, setting a new SOTA for this repair
setting and outperforming model-trusting, robust, tuned-robust, and chance-constrained
repair on the deployable feasibility-cost frontier. More broadly, TCR is a step toward trustworthy,
physics-aware AI for engineering decisions whose models are accurate almost everywhere but
wrong exactly where it matters; the same discover-calibrate-repair-attribute principle
applies wherever learned or optimized decisions must be deployed against an imperfectly
known physical world.

\bibliography{Wang}

\end{document}